\documentclass[aps,prl,twocolumn,showpacs,superscriptaddress]{revtex4} 

\usepackage{amsmath,amssymb,mathrsfs}
\usepackage{latexsym}
\usepackage{graphicx} 
\usepackage{epstopdf}
\usepackage{graphicx,epstopdf,color}
\usepackage{amsfonts}
\usepackage{bm}

\newcommand{\bs}[1]{\boldsymbol{#1}}

\def\ie{\emph{i.e.},\ }

\begin{document}

\title{Spiral order in the honeycomb iridate Li$_2$IrO$_3$}

\author{Johannes Reuther}
\affiliation{Department of Physics, California Institute of Technology, Pasadena, CA 91125, USA}
\author{Ronny Thomale}
\affiliation{Institute for Theoretical Physics, University of W\"urzburg, Am Hubland, D-97074 W\"urzburg, Germany}
\author{Stephan Rachel}
\affiliation{Institut f\"ur Theoretische Physik, Technische Universit\"at Dresden, 01062 Dresden, Germany}

 \pagestyle{plain}

\begin{abstract}
The honeycomb iridates A$_2$IrO$_3$ (A=Na, Li) constitute promising candidate materials to realize the Heisenberg-Kitaev model (HKM) in nature, hosting unconventional magnetic as well as spin liquid phases.
Recent experiments suggest, however, that Li$_2$IrO$_3$ exhibits a magnetically ordered state of incommensurate spiral type which has not been identified in the HKM. We show that these findings can be understood in the context of an extended Heisenberg-Kitaev scenario satisfying all tentative experimental evidence: (i) the maximum of the magnetic susceptibility is located inside the first Brillouin zone, (ii) the Curie-Weiss temperature is negative relating to dominant antiferromagnetic fluctuations, and (iii) significant second-neighbor spin-exchange is involved.
\end{abstract}


\maketitle

%
%

\begin{figure}[t]
\centering
\includegraphics[width=0.8\linewidth]{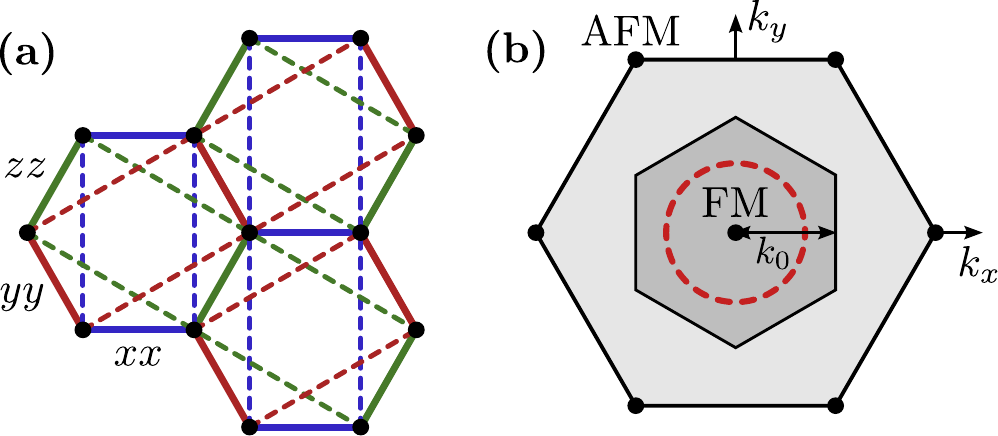}
\caption{(a) Different colors of the nearest (full lines) and next nearest (dashed lines) neighbor bonds on the honeycomb lattice represent Kitaev interactions of $S_i^x S_j^x$-type (blue), $S_i^y S_j^y$-type (red) and $S_i^z S_j^z$-type (green). (b) Extended Brillouin zone scheme (inner hexagon is the first Brillouin zone) of the 
honeycomb lattice. Ferromagnetic (FM) order manifests as peaks in the center of the first Brillouin zone, while antiferromagnetic (AFM) order resides at the corner of the extended zone scheme. The spiral order found in experiments corresponds to an ordering wave vector on the red ring well inside the first Brillouin zone. 
$k_0$ denotes the distance from the $\Gamma$-point to the first Brillouin zone boundary.\\[5pt]}
\label{fig:honeycomb}
\end{figure}

{\it Introduction}.---Transition metal oxides such as Iridates have attracted considerable attention recently. The interest is especially driven by the intriguing interplay of strong spin-orbit coupling and electronic correlations, potentially leading to unconventional quantum magnetism or paramagnetism such as spin liquids. The iridium oxides A$_2$IrO$_3$ (A=Na, Li) have caused particular excitement since it has been suggested that they realize the Heisenberg-Kitaev model (HKM)\,\cite{jackeli-09prl017205,chaloupka-10prl027204} on the honeycomb lattice (Fig.~\ref{fig:honeycomb}a). The Kitaev limit of this model provides a platform for a spin liquid with fractional anyonic excitations\,\cite{kitaev06ap2}. 
A vivid debate has been triggered on the suitable microscopic model describing honeycomb iridates as well as their experimental signatures\,\cite{shitade-09prl256403,jackeli-09prl017205,chaloupka-10prl027204,reuther-11prb100406,singh-12prl127203,mazin-12prl197201,price-12prl187201,kim-12prl106401,comin-12prl266406,cao-13prb220414,gretarsson-13prl076402,gretarsson-13prb220407,foyevtsova-13prb035107,jenderka-13prb045111,andrade-13arXiv:1309.2951,manni-13arXiv:1312.0815,trousselet-13prl037205,kim-14prb081109,rau-14prl077204,katukuri-13arXiv:1312.7437,nishimoto-14arXiv:1403.6698}, and whether there is some material located in or in proximity to the Kitaev spin liquid.

So far, most experiments have focussed on the sodium compound\,\cite{singh-82prb064412} which 
turned out to exhibit zigzag magnetic order instead of being a spin liquid\,\cite{liu-11prb220403,choi-12prl127204,ye-12prb180403}. This finding was rather unexpected since the HKM as originally proposed \,\cite{jackeli-09prl017205,chaloupka-10prl027204} does not host a zigzag ordered phase. Several extensions of the HKM such as significant longer range Heisenberg interactions have been discussed in order to possibly explain the occurrence of this type of order\,\cite{chaloupka-13prl097204,kimchi-11prb180407,singh-12prl127203,mazin-12prl197201}.

Recent experiments have investigated the lithium compound and found magnetic long-range order below $T_N=15$\,K\,\cite{singh-12prl127203}. Smaller trigonal distortions of the IrO$_6$ octahedra due to the enhanced electro-negativity of Li might lead to stronger Kitaev-like interactions. It  has further been
suggested that the magnetic order is different as compared to the Na compound\,\cite{cao-13prb220414,manni-13arXiv:1312.0815}.
Latest neutron scattering experiments revealed that the magnetic order is of incommensurate spiral type \cite{coldea13}. 
Using neutron powder diffraction, it was observed that the absolute value of the magnetic Bragg peak resides inside the first Brillouin zone (red dashed line in Fig.\,\ref{fig:honeycomb}b)\,\cite{coldea13}. 
Most recently, the depletion of Li$_2$IrO$_3$ with non-magnetic Ti-atoms\,\cite{gegenwart-unpub} was shown to result in a characteristic behavior of the spin-glass temperature\,\cite{andrade-13arXiv:1309.2951}. This suggests that spin exchange beyond nearest-neighbors is dominating.

This result is even more puzzling than the findings for Na$_2$IrO$_3$: Firstly, the HKM which is believed to describe the iridates does not contain a spiral ordered phase. As shown below, the canonical extension via longer range Heisenberg couplings will not be sufficient to account for the experimental evidence.
Secondly, the small wave vector of the tentative magnetic order in Li$_2$IrO$_3$ necessitates a 
spin model exhibiting the astonishing coincidence of pronounced ferromagnetic interactions along with a negative Curie-Weiss-temperature ($-33$K)\,\cite{singh-12prl127203} hinting at dominant antiferromagnetic fluctuations. Thirdly, significant second-neighbor spin-exchange must be involved.

In this letter, we show that the Heisenberg-Kitaev model extended by next-nearest neighbor Heisenberg {\it and} Kitaev interactions is capable of describing the experimental evidence of magnetism in Li$_2$IrO$_3$: This model realizes the spiral order observed, and allows us to devise a mechanism to reconcile the joint occurrence of magnetic order at small wave vectors and an antiferromagnetic Curie-Weiss temperature along with significant second-neighbor spin-exchange.

\begin{figure}[t]
\centering
\includegraphics[width=0.9\linewidth]{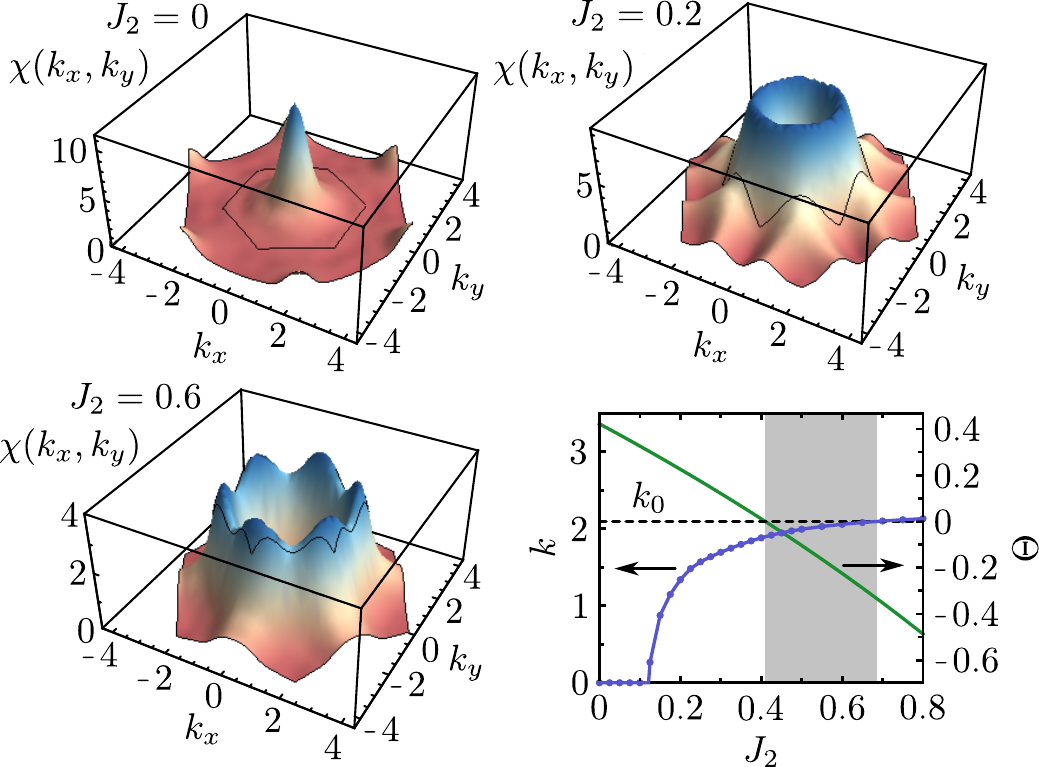}
\caption{Susceptibility profiles for the $J_1$-$J_2$ Heisenberg model Eq.~(\ref{ferro}) and $J_1=-1$. Thin black lines mark the boundary of the first Brillouin zone part within the extended Brillouin zone. For small $J_2>0$ we first detect FM order. Above $J_2\approx 0.12$ the peaks split resulting in incommensurate spiral peaks, see main text for explanations.
Bottom right: Peak position $k=|{\mathbf k}|$ and Curie-Weiss temperature $\Theta$ as a function of $J_2$. $k_0=2\pi/3$ is defined in Fig.\,\ref{fig:honeycomb}b. The gray shaded region is the parameter regime with spiral peaks inside the first Brillouin zone and negative Curie-Weiss temperature.}
\label{fig:sus_profiles_j2}
\end{figure}
%

%
%

{\it $J_1<0$ Heisenberg coupling}.---A straightforward way to realize spiral order inside the first Brillouin zone is given by the isotropic $J_1$-$J_2$-Heisenberg model on the honeycomb lattice
\begin{equation}
H=J_1\sum_{\langle ij\rangle}{\mathbf S}_i {\mathbf S}_j+J_2\sum_{\langle\langle ij\rangle\rangle}{\mathbf S}_i {\mathbf S}_j\label{ferro}
\end{equation}
with $J_1<0$ and $J_2>0$. We have investigated this model using the functional renormalization-group technique based on pseudo fermions (PFFRG) which includes quantum fluctuations beyond RPA or spin-wave theory and which has been successfully applied to various honeycomb systems\,\cite{reuther-10prb144410,PhysRevB.84.014417,reuther-11prb100406,reuther-12prb155127}; details of the method are provided in the supplemental material\,\cite{sm}. As shown in Fig.\,\ref{fig:sus_profiles_j2} (top left) for $J_2=0$, the susceptibility shows a sharp FM peak in the center of the Brillouin zone. Switching on $J_2$, this peak first broadens and, above $J_2\approx0.12$, forms a ring at incommensurate spiral wave vectors with increasing diameter for larger $J_2$ (see Fig.\,\ref{fig:sus_profiles_j2}). In particular around $J_2=0.2$, such profiles resemble the experimental findings of spiral magnetic order inside the first Brillouin zone. We argue, however, that this scenario of interactions is unlikely: plotting the peak positions $k=|{\mathbf k}|$ 
together with the Curie-Weiss temperatures $\Theta$ (from a fit $\chi(k=0,T)\sim 1/(T-\Theta)$ of our susceptibility data\,\cite{sm}) shows that there is indeed a parameter regime $0.4\lesssim J_2\lesssim0.7$ where the susceptibly maximum is inside the first Brillouin zone and $\Theta$ is negative, see Fig.\,\ref{fig:sus_profiles_j2} (bottom right). However, in this regime the peaks are very close to the edges of the first Brillouin zone, in disagreement with experimental results. More importantly, the PFFRG detects very strong quantum fluctuations for such parameters, indicating the suppression of any magnetic order beyond what is found experimentally\,\cite{sm,footnote-PFFRG}.
\begin{figure}[t]
\centering
\includegraphics[width=0.6\linewidth]{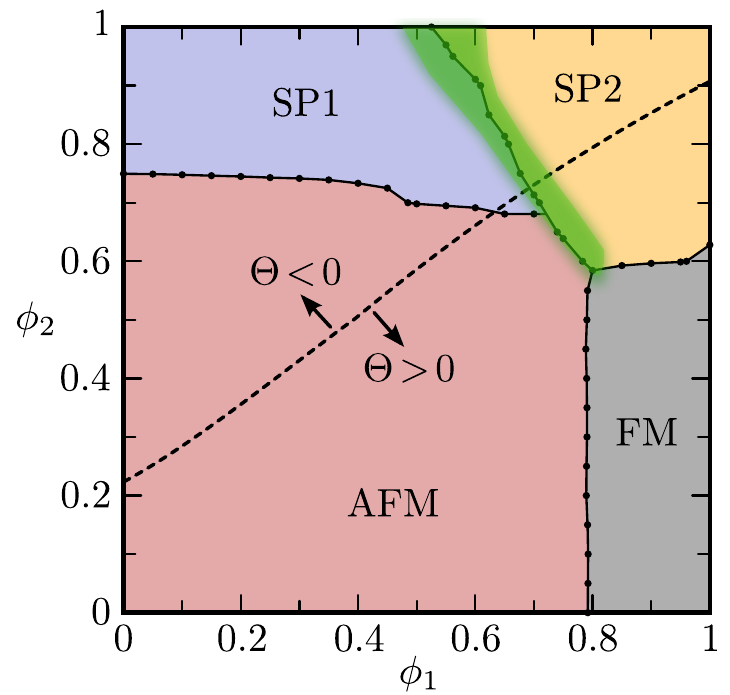}
\caption{Phase diagram of the extended HKM in Eq.\,(\ref{kitaev}), $g=0.8$. We find FM order, AFM order, incommensurate spiral order with wave vectors outside the first Brillouin zone (SP1), and incommensurate spiral order with wave vectors inside the first Brillouin zone (SP2). Shaded areas indicate enhanced quantum fluctuations, possibly signaling a narrow non-magnetic phase. The dashed line separates parameter regimes with positive from negative Curie-Weiss temperature.}
\label{fig:phase_diag_phi1_phi2}
\end{figure}
\begin{figure*}[t]
\centering
\includegraphics[width=0.95\linewidth]{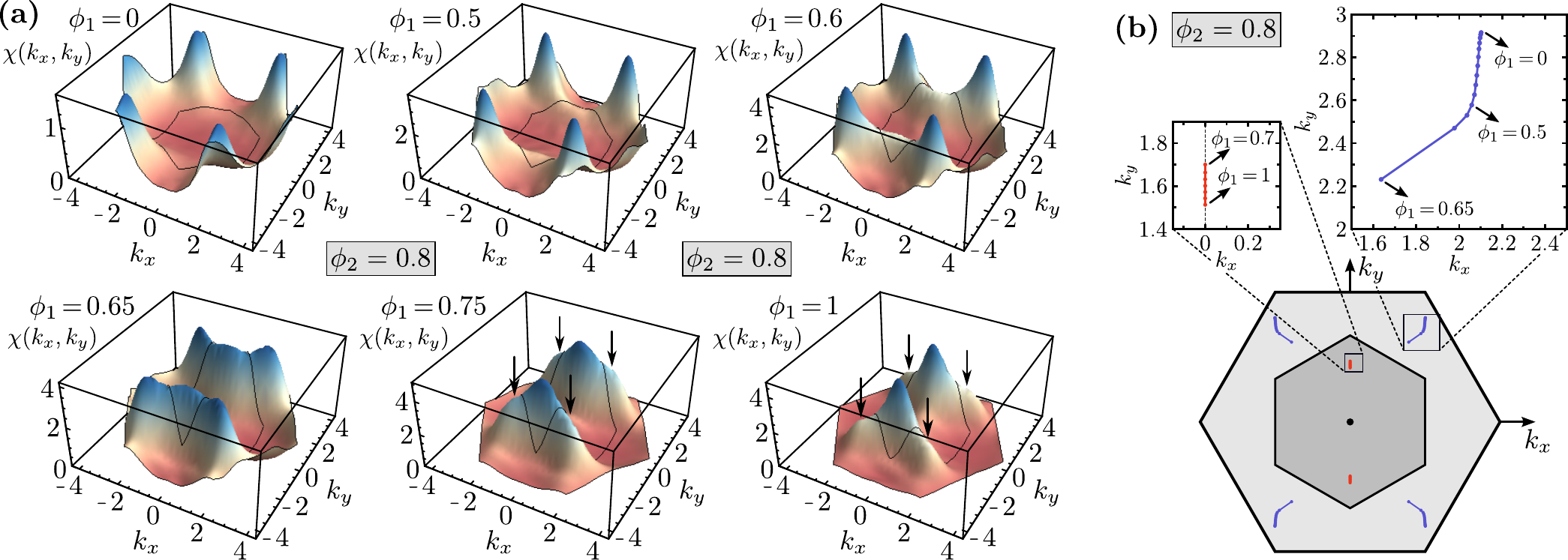}
\caption{(a) Susceptibility profiles for the spiral phases of Fig.\,\ref{fig:phase_diag_phi1_phi2}, along a cut with $\phi_2=0.8$ and Brillouin zone notation as in Fig.\,\ref{fig:sus_profiles_j2}. For larger $\phi_1$, new ordering peaks emerge in the first Brillouin zone (SP2 order). Residual SP1 signatures persist, manifesting via shoulders marked by arrows. All plots display the $xx$-component of susceptibility. The corresponding $yy$- and $zz$- components result from $2\pi/3$ rotations in ${\mathbf k}$ space. (b) Detailed migration profile of the ordering peaks (blue, red).}
\label{fig:sus_profiles_phi1}
\end{figure*}
We emphasize that deviating signs of $J_1$, $J_2$ and/or additional third neighbor exchange $J_3$ as well as FM nearest-neighbor Kitaev couplings (Fig.\,\ref{fig:honeycomb}a) do not change our conclusion: never do we find a magnetically ordered regime with spiral peaks deep inside the first Brillouin zone, combined with a negative Curie-Weiss temperature. For generic spin models on the honeycomb lattice, the susceptibility peak position at the edge of the first Brillouin zone approximately corresponds to the boundary between positive and negative Curie-Weiss temperatures.

%
%

{\it Second neighbor Kitaev exchange}.---We now consider AFM nearest-neighbor Heisenberg exchange $J_1>0$ and FM nearest-neighbor Kitaev exchange $K_1<0$ as originally proposed for the HKM\,\cite{jackeli-09prl017205,chaloupka-10prl027204}. Substantiated by {\it ab initio} calculations, such signs of interactions seem to be most likely\,\cite{mazin-12prl197201,foyevtsova-13prb035107,katukuri-13arXiv:1312.7437}. Furthermore, we consider FM isotropic second-neighbor exchange $J_2<0$ and AFM second-neighbor Kitaev couplings $K_2>0$ (for the convention of $K_2$ couplings, see Fig.\,\ref{fig:honeycomb}a). {\it It turns out that $K_2$ couplings are of great importance for our considerations and represent the crucial step towards an understanding of the experimental results.}
Such longer-ranged Kitaev terms have originally been deduced from a strong coupling expansion of the band structure for Na$_2$IrO$_3$\,\cite{shitade-09prl256403,reuther-12prb155127}. Second-neighbor Kitaev exchange $K_2$ stems from spin-orbit coupling, which is likely to play a dominant role 
for the electronic state of iridates (see, e.g., \cite{mazin-12prl197201,foyevtsova-13prb035107}). 

As argued in Ref.\,\cite{jackeli-09prl017205}, the IrO$_6$ octahedra in A$_2$IrO$_3$ share their edges leading to two 90$^\circ$ Ir-O-Ir exchange paths; projection onto the lowest Kramers doublet
results in FM nearest neighbor Kitaev interactions $K_1<0$. In addition, direct overlap of Ir orbitals on neighboring sites leads to ordinary AFM nearest neighbor Heisenberg exchange with $J_1>0$. We also consider longer-ranged hopping processes with real and imaginary transfer integrals\,\cite{shitade-09prl256403,mazin-12prl197201,foyevtsova-13prb035107}. 
In the Mott limit, these bond-selective spin-orbit hoppings correspond to a $J'>0$ second neighbor coupling \,\cite{reuther-12prb155127,khaliullin05ptps155}:
\begin{equation}\nonumber
H_\text{NNN}  = \sum_{\langle\!\langle ij \rangle\!\rangle_\gamma} J' \big[\, 2 S_i^\gamma S_j^\gamma -  \bs{S}_i \bs{S}_j\,\big]\ .
\end{equation}
We see that aside from an AFM Kitaev term, the spin-orbit coupling also generates second-neighbor FM Heisenberg exchange. In addition, we allow for small deviations in the isotropic Heisenberg exchange by including real second-neighbor hopping resulting in AFM spin exchange with amplitude $J_0'>0$. 
The total second neighbor spin Hamiltonian then reads
$
H_{\rm NNN} = \sum_{\langle\!\langle ij \rangle\!\rangle_\gamma} 2J' S_i^\gamma S_j^\gamma +( J_0' - J') \bs{S}_i \bs{S}_j$.
As we consider the real second neighbor hoppings to be small compared to the imaginary ones, we assume $J_0'-J'<0$.  Setting $2J'\equiv K_2$ and $J_0'-J'\equiv J_2$, we obtain
\begin{equation}\begin{split}
H~=~&J_1\sum_{\langle ij\rangle}{\mathbf S}_i{\mathbf S}_j+K_1\sum_{\langle ij\rangle_\gamma}S^\gamma_i S^\gamma_j \\
+&J_2\sum_{\langle\!\langle ij\rangle\!\rangle}{\mathbf S}_i{\mathbf S}_j+K_2\sum_{\langle\!\langle ij\rangle\!\rangle_\gamma}S^\gamma_i S^\gamma_j\,,\label{kitaev}
\end{split}\end{equation}
where $\gamma$ denotes the bond-selective anisotropies as shown in Fig.~\ref{fig:honeycomb}a. Eq.\,\eqref{kitaev} is what we believe to be the minimal model for magnetism in the honeycomb iridates. We parametrize the different couplings as $J_1=\cos(\pi\phi_1/2)$, $K_1=-\sin(\pi\phi_1/2)$, $J_2=-g\cos(\pi\phi_2/2)$, $K_2=g\sin(\pi\phi_2/2)$ with $\phi_{1,2}\in[0,1]$ and $g\geq  0$. $\phi_{1(2)}$ changes the relative strength of Heisenberg and Kitaev interactions for (next) nearest neighbor couplings. Furthermore, $g$ is the total relative strength of first and second neighbor exchange. Note that $J_0'=0$ corresponds to $\phi_2\approx 0.7$, as considered in Ref.\,\onlinecite{reuther-12prb155127}.

We have performed extensive calculations on Eq.~\,\eqref{kitaev} via PFFRG. Within a wide range of $g$, \ie $0.4\lesssim g\lesssim2$ the phase diagram is approximately constant. As a representative case, we consider $g=0.8$ in the following. The resulting phase diagram as a function of $\phi_1\in[0,1]$ and $\phi_2\in[0,1]$ is shown in Fig.\,\ref{fig:phase_diag_phi1_phi2}. We find four magnetically ordered phases: FM order, AFM order, incommensurate spiral order with wave vectors outside the first Brillouin zone (SP1) and incommensurate spiral order with wave vectors inside the first Brillouin zone (SP2). It can be seen that for prominent $K_2$, there is an extended SP2-phase with negative Curie-Weiss temperature $\Theta$. We note that the origin of spiral phases for a similar model has been discussed in Ref.\,\onlinecite{reuther-12prb155127}.

Fig.\,\ref{fig:sus_profiles_phi1}a shows susceptibility profiles along the cut $\phi_2=0.8$. In the SP1 phase (addressed in Refs.\,\cite{reuther-12prb155127,liu-13prb245119,kargarian-12prb205124}) at small $\phi_1$, there are four ordering peaks located outside the first Brillouin zone. As $\phi_1$ increases, the ferromagnetic interactions become stronger such that the ordering peaks move towards the $\Gamma$-point. At $\phi_1\approx0.65$ new peaks inside the first Brillouin zone emerge, and the overall maxima jump to these new positions indicating the onset of the SP2 phase. Increasing $\phi_1$ the two remaining ordering peaks further move inside. In the SP2 phase, there are persistent sub-leading signatures (``shoulders'' marked by arrows in Fig.\,\ref{fig:sus_profiles_phi1}a) inherited from the SP1 peaks. A migration profile of the ordering peaks is depicted in Fig.\,\ref{fig:sus_profiles_phi1}b.

\begin{figure}[t]
\centering
\includegraphics[width=0.99\linewidth]{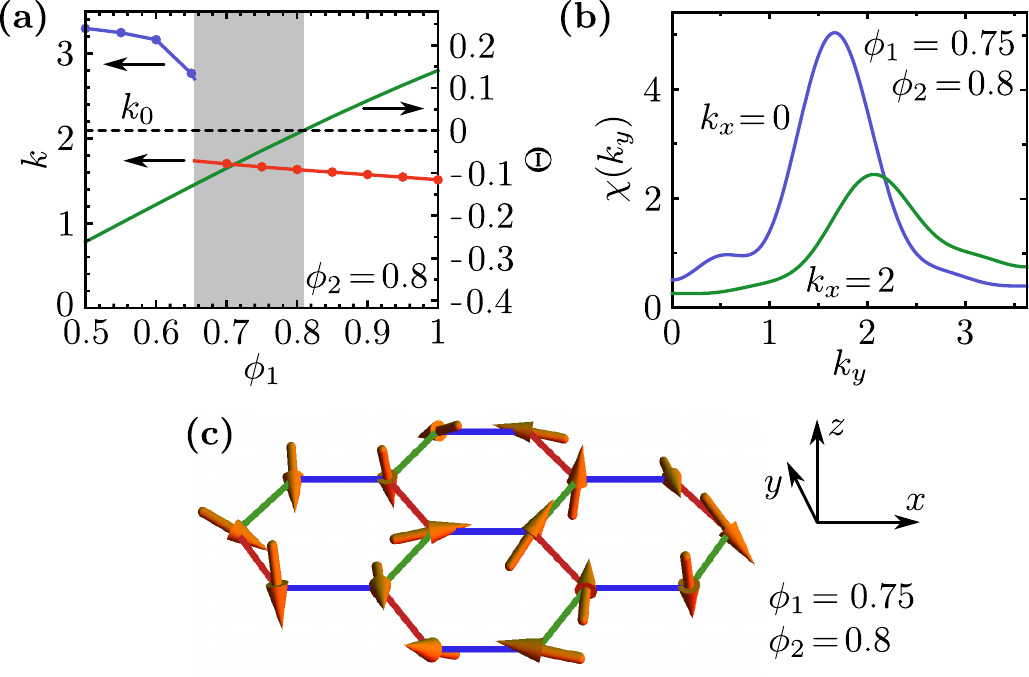}
\caption{(a) Absolute value $k$ of the wave vector at the ordering peak and the Curie-Weiss temperature in the SP1 phase (blue) and in the SP2 phase (red) as a function of $\phi_1$ ($\phi_2=0.8$). The jump in the peak position for $k$ is clearly observed. The gray shaded region marks the joint appearance of spiral peaks inside the first Brillouin zone and negative Curie-Weiss temperatures. 
(b) Cut through the susceptibility at  $k_x=0$ (blue) and $k_x=2$ (green) as a function of $k_y$. The Bragg-peak maximum is at $\bs{k}=(0,1.66)/a_{\rm Ir-Ir}$. (c) The spin pattern related to Li$_2$IrO$_3$ forms a nonplanar spiral.}
\label{fig:k_lambdacw}
\end{figure}

The SP2 phase is characterized by ordering peaks located well inside the first Brillouin zone which can occur along with a negative $\Theta$. This is illustrated in Fig.\,\ref{fig:k_lambdacw}a displaying the absolute value $k$ of the ordering peak and the Curie-Weiss temperature as a function of $\phi_1$ at constant $\phi_2=0.8$. 
The magnetic profile in this parameter regime is, hence, in agreement with the experimental results, suggesting that the extended HKM of Eq.\,\eqref{kitaev} provides a suitable description of Li$_2$IrO$_3$. 
From Fig.\,\ref{fig:sus_profiles_phi1}a it is also clear why an SP2 phase with negative Curie-Weiss temperature is possible: SP2 still exhibits sub-leading ordering tendencies with wave vectors outside the first Brillouin zone, which manifest as the aforementioned shoulders in the susceptibility profiles. While these antiferromagnetic-type ordering fluctuations do not yield long-range magnetic order, they still shift the Curie-Weiss temperature towards negative values. The special properties of this parameter regime crucially rely on a strong $K_2$ exchange, as we could not find a similar phenomenology without $K_2$. As $K_2$ stems from spin-orbit coupling our findings are in agreement with the commonly accepted picture that spin-orbit coupling plays a dominant role in the honeycomb iridates\,\cite{jackeli-09prl017205,shitade-09prl256403,foyevtsova-13prb035107}.

Fig.\,\ref{fig:k_lambdacw}b shows different cuts displaying significant weight for larger $k$ which is responsible for the negative Curie-Weiss temperature.
We therefore predict that susceptibility enhancements outside the first Brillouin zone should be visible upon probing this domain for Li$_2$IrO$_3$. 
In Fig.\,\ref{fig:k_lambdacw}c we illustrate the classical spin pattern corresponding to the quantum magnetic order in the SP2 phase. Different types of incommensurate spiral orders on the honeycomb lattice are classified according to their symmetry properties. The location of ordering peaks in $k$-space indicates that the spiral in the SP2 phase is of so-called H1-type\,\cite{Rastelli13,footnote3}. The intrinsic relation between real space and spin  space transformations in the Kitaev model further requires that the $x$-, $y-$, and $z-$components of the real space spin-spin correlation function are rotated by $120^\circ$ among each other. By enforcing this condition one finds a nonplanar spiral as shown in Fig.\,\ref{fig:k_lambdacw}c.

It is worth mentioning that the qualitative features of the SP2 phase persist when we reduce $g$ (\ie the ratio between nearest and second-nearest neighbor interactions), until at small enough $g$ the Kitaev spin liquid sets in. Hence, depending on the precise value of $g$ hypothetically realized in Li$_2$IrO$_3$ (which we cannot determine within the present analysis), the compound might be located in close vicinity to a Kitaev spin liquid phase. Note that the pure $K_1$--$K_2$ model already hosts both the Kitaev spin liquid and the SP2 phase, although the quantitative features of the SP2 phase found therein do not agree with experiment.

%
%

{\it Conclusion}.---We have shown that the Heisenberg-Kitaev model extended to next nearest neighbor Heisenberg and Kitaev couplings emerges as a promising minimal model to explain the puzzling situation for the magnetic profile of Li$_2$IrO$_3$: in the experimentally relevant parameter regime proposed by us, (i) the magnetic order is of incommensurate spiral type with ordering peaks  located well inside the first Brillouin zone, (ii) the Curie-Weiss temperature is negative, and (iii) significant second-neighbor spin-exchange is involved ($g=0.8$). We claim that the simultaneous fulfillment of (i) and (ii) is connected to sub-leading susceptibility peaks outside the first Brillouin zone which establish a promising line of investigation for future experiments.


\begin{acknowledgements}
We thank R.\ Coldea for insightful comments on the manuscript and 
acknowledge discussions with R.\ Valenti, I.\ Mazin, D.\ Inosov, G.\ Khaliullin, P.\ Gegenwart, S.\ Manni, E.\ Andrade, and M.\ Vojta.
JR acknowledges support by the Deutsche Akademie der Naturforscher Leopoldina through grant LPDS 2011-14.
RT is supported by the ERC starting grant TOPOLECTRICS of the European Research Council (ERC-StG-2013-336012).
SR is supported by the DFG priority program SPP 1666 ``Topological Insulators'', DFG FOR 960, and by the Helmholtz association through VI-521. 
\end{acknowledgements}

\bibliographystyle{prsty}
\bibliography{lithium_iridate}

\begin{thebibliography}{10}

\bibitem{jackeli-09prl017205}
G. Jackeli and G. Khaliullin, Phys. Rev. Lett. {\bf 102},  017205  (2009).

\bibitem{chaloupka-10prl027204}
J. Chaloupka, G. Jackeli, and G. Khaliullin, Phys. Rev. Lett. {\bf 105},
  027204  (2010).

\bibitem{kitaev06ap2}
A.~Y. Kitaev, Ann. Phys. (N.Y.) {\bf 321},  2  (2006).

\bibitem{shitade-09prl256403}
A. Shitade, H. Katsura, J. Kune\ifmmode~\check{s}\else \v{s}\fi{}, X.-L. Qi,
  S.-C. Zhang, and N. Nagaosa, Phys. Rev. Lett. {\bf 102},  256403  (2009).

\bibitem{reuther-11prb100406}
J. Reuther, R. Thomale, and S. Trebst, Phys. Rev. B {\bf 84},  100406  (2011).

\bibitem{singh-12prl127203}
Y. Singh, S. Manni, J. Reuther, T. Berlijn, R. Thomale, W. Ku, S. Trebst, and
  P. Gegenwart, Phys. Rev. Lett. {\bf 108},  127203  (2012).

\bibitem{mazin-12prl197201}
I.~I. Mazin, H.~O. Jeschke, K. Foyevtsova, R. Valent\'\i, and D.~I. Khomskii,
  Phys. Rev. Lett. {\bf 109},  197201  (2012).

\bibitem{price-12prl187201}
C.~C. Price and N.~B. Perkins, Phys. Rev. Lett. {\bf 109},  187201  (2012).

\bibitem{kim-12prl106401}
C.~H. Kim, H.~S. Kim, H. Jeong, H. Jin, and J. Yu, Phys. Rev. Lett. {\bf 108},
  106401  (2012).

\bibitem{comin-12prl266406}
R. Comin {\it et~al.}, Phys. Rev. Lett. {\bf 109},  266406  (2012).

\bibitem{cao-13prb220414}
G. Cao, T.~F. Qi, L. Li, J. Terzic, V.~S. Cao, S.~J. Yuan, M. Tovar, G. Murthy,
  and R.~K. Kaul, Phys. Rev. B {\bf 88},  220414(R)  (2013).

\bibitem{gretarsson-13prl076402}
H. Gretarsson {\it et~al.}, Phys. Rev. Lett. {\bf 110},  076402  (2013).

\bibitem{gretarsson-13prb220407}
H. Gretarsson {\it et~al.}, Phys. Rev. B {\bf 87},  220407(R)  (2013).

\bibitem{foyevtsova-13prb035107}
K. Foyevtsova, H.~O. Jeschke, I.~I. Mazin, D.~I. Khomskii, and R. Valent\'\i,
  Phys. Rev. B {\bf 88},  035107  (2013).

\bibitem{jenderka-13prb045111}
M. Jenderka, J. Barzola-Quiquia, Z. Zhang, H. Frenzel, M. Grundmann, and M.
  Lorenz, Phys. Rev. B {\bf 88},  045111  (2013).

\bibitem{andrade-13arXiv:1309.2951}
E.~C. Andrade and M. Vojta, arXiv:1309.2951.

\bibitem{manni-13arXiv:1312.0815}
S. Manni, S. Choi, I.~I. Mazin, R. Coldea, M. Altmeyer, H.~O. Jeschke, R.
  Valenti, and P. Gegenwart, arXiv:1312.0815.

\bibitem{trousselet-13prl037205}
F. Trousselet, M. Berciu, A.~M. Ole\ifmmode~\acute{s}\else \'{s}\fi{}, and P.
  Horsch, Phys. Rev. Lett. {\bf 111},  037205  (2013).

\bibitem{kim-14prb081109}
B.~H. Kim, G. Khaliullin, and B.~I. Min, Phys. Rev. B {\bf 89},  081109
  (2014).

\bibitem{rau-14prl077204}
J.~G. Rau, E.~K.-H. Lee, and H.-Y. Kee, Phys. Rev. Lett. {\bf 112},  077204
  (2014).

\bibitem{katukuri-13arXiv:1312.7437}
V.~M. Katukuri, S. Nishimoto, V. Yushankhai, A. Stoyanova, H. Kandpal, S. Choi,
  R. Coldea, I. Rousochatzakis, L. Hozoi, and J. van~den Brink,
  arXiv:1312.7437.

\bibitem{nishimoto-14arXiv:1403.6698}
S. Nishimoto, V.~M. Katukuri, V. Yushankhai, H. Stoll, U.~K. Roessler, L.
  Hozoi, I. Rousochatzakis, and J. {van den Brink}, arXiv:1403.6698.

\bibitem{singh-82prb064412}
Y. Singh and P. Gegenwart, Phys. Rev. B {\bf 82},  064412  (2010).

\bibitem{liu-11prb220403}
X. Liu, T. Berlijn, W.-G. Yin, W. Ku, A. Tsvelik, Y.-J. Kim, H. Gretarsson, Y.
  Singh, P. Gegenwart, and J.~P. Hill, Phys. Rev. B {\bf 83},  220403(R)
  (2011).

\bibitem{choi-12prl127204}
S.~K. Choi {\it et~al.}, Phys. Rev. Lett. {\bf 108},  127204  (2012).

\bibitem{ye-12prb180403}
F. Ye, S. Chi, H. Cao, B.~C. Chakoumakos, J.~A. Fernandez-Baca, R. Custelcean,
  T.~F. Qi, O.~B. Korneta, and G. Cao, Phys. Rev. B {\bf 85},  180403(R)
  (2012).

\bibitem{chaloupka-13prl097204}
J. Chaloupka, G. Jackeli, and G. Khaliullin, Phys. Rev. Lett. {\bf 110},
  097204  (2013).

\bibitem{kimchi-11prb180407}
I. Kimchi and Y.-Z. You, Phys. Rev. B {\bf 84},  180407  (2011).

\bibitem{coldea13}
{R. Coldea}, {T}alk at the SPORE13 conference held at MPI-PKS, Dresden (2013);
  {S. Choi}, {T}alk at the APS March meeting, Denver, CO (2014); {R. Coldea},
  {T}alk at the DPG-focus session, spring meeting of the german physical
  society, Dresden (2014).

\bibitem{gegenwart-unpub}
S. Manni, Y. Tokiwa, and P. Gegenwart, Phys. Rev. B {\bf 89},  241102(R)
  (2014).

\bibitem{reuther-10prb144410}
J. Reuther and P. W\"olfle, Phys. Rev. B {\bf 81},  144410  (2010).

\bibitem{PhysRevB.84.014417}
J. Reuther, D.~A. Abanin, and R. Thomale, Phys. Rev. B {\bf 84},  014417
  (2011).

\bibitem{reuther-12prb155127}
J. Reuther, R. Thomale, and S. Rachel, Phys. Rev. B {\bf 86},  155127  (2012).

\bibitem{sm}
{S}ee {S}upplemental {M}aterial at http://link.aps.org/
  supplemental/xxxx/PhysRevLett.xxxx.

\bibitem{footnote-PFFRG}
The absence of magnetic order is not surprising as our parameters correspond to
  a critical line in the classical phase diagram\,\cite{Rastelli13}.

\bibitem{khaliullin05ptps155}
G. Khaliullin, Prog. Theor. Phys. Suppl. {\bf 160},  155  (2005).

\bibitem{liu-13prb245119}
T. Liu, B. Dou\ifmmode~\mbox{\c{c}}\else \c{c}\fi{}ot, and K. Le~Hur, Phys.
  Rev. B {\bf 88},  245119  (2013).

\bibitem{kargarian-12prb205124}
M. Kargarian, A. Langari, and G.~A. Fiete, Phys. Rev. B {\bf 86},  205124
  (2012).

\bibitem{Rastelli13}
E. Rastelli, {\em Statistical Mechanics of Magnetic Excitations} (World
  Scientific, Singapore, 2013).

\bibitem{footnote3}
The H1 phase is defined so that the two inequivalent spins in the unit cell are
  parallel. In contrast, in the H2 phase a relative angle $\alpha$ between the
  two spins in the unit cell is allowed\,\cite{Rastelli13}.

\end{thebibliography}

\end{document}